# Thoughts on opportunities from high-energy nuclear collisions

*Federico Antinori, Nestor Armesto, Paolo Bartalini, René Bellwied, Peter Braun-Munzinger, Brian Cole, Andrea Dainese, Marek Gazdzicki, Paolo Giubellino, John Harris, Ulrich Heinz, Barbara Jacak, Peter Jacobs, Dmitri Kharzeev, Constantin Loizides, Silvia Masciocchi, Andreas Morsch, Berndt Mueller, Jamie Nagle, Guy Paic, Krishna Rajagopal, Gunther Roland, Karel Safarik, Jurgen Schukraft, Yves Schutz, Johanna Stachel, Peter Steinberg, Thomas Ullrich, Xin-Nian Wang, Johannes Wessels, Urs Achim Wiedemann.*

## Abstract

*This document summarizes thoughts on opportunities from high-energy nuclear collisions.*

## Introduction

On the weekend of 24-26 May 2014, a group of 31 physicists met in the monastery Mont Sainte Odile in the French Vosges to discuss the current status of high-energy nuclear collisions and the scientific opportunities arising from it. The meeting was centered around the questions: What are the qualitatively most interesting features in existing data that warrant further study and how do they lend themselves to further experimentation? What are the central questions about the properties of hot and dense nuclear matter that come in reach by addressing them? And which next-generation developments are needed on the experimental and theoretical side to exploit them?

All participants were asked prior to the meeting to formulate their main considerations related to these questions in the form of a few itemized statements. During the meeting, participants introduced their arguments in short 10-minute interventions that were followed by open discussion. In this way, the usual theory and experimental talks were substituted by structured but free-floating exchanges on the most burning issues. In this non-traditional setting, the discussion focused quickly on a small number of central topics that – due to the broadly complementary experience of the participants present – could be discussed from different perspectives. The present document summarizes the main topics identified in the discussions on Mont Sainte Odile, together with those itemized conclusions that emerged as a **consensus view** amongst the participants.

The participants met as individual scientists interested in the study of extreme forms of QCD matter via high energy nuclear collisions. They neither met with any mandate from a collaboration, nor from their scientific community nor from funding agencies. They also met without any preconceived idea that a document must be written. However, they met with the firm belief that with the data available by now from A-A, p/d-A and pp collisions at the LHC and at RHIC, it is time to discuss on how to follow up on the discoveries of recent years and on the currently prevailing interpretational paradigm with a programme of detailed experimentation and focused theoretical exploration.

The participants believe that the points formulated in this short document can be useful in this much-needed process and it is for this reason that these conclusions are made available via this posting on the arXiv.

Readers interested in raising constructive critique, communicating complementary viewpoints or taking the points listed in this write-up further, are invited to leave critical comments on the twiki https://twiki.cern.ch/twiki/bin/view/MontSainteOdileThinkTank/WorldComments2

## Initial State

The wave functions of hadrons and nuclei are of fundamental interest. The decoherence of these wave functions in relativistic heavy-ion collisions and the evolution of this initial state towards a strongly-interacting fluid are as yet poorly constrained, and offer an exciting window into the physics of equilibration in QCD.

- Our present understanding of the processes transforming the initial quantum-state of matter in a relativistic heavy-ion collision into a hydrodynamic fluid is substantially incomplete. Weakly coupled calculations connect smoothly to the initial quantum state if that state is weakly coupled, but have difficulty connecting to hydrodynamics. Strongly coupled calculations yield hydrodynamic fluids smoothly and automatically, but assume a strongly coupled initial quantum state. It is therefore essential to develop experimental probes that can determine whether the gluons in the initial state wave function are weakly or strongly coupled.

- Holographic methods that use dualities to map difficult dynamical questions in a strongly coupled gauge theory to tractable classical gravity calculations have taught us over the last decade numerous qualitative lessons about strongly coupled plasma, heavy quarks therein, and jet quenching. Rapid formation of a hydrodynamic fluid is found in collisions of sheets or disks of strongly coupled cold quantum matter. Progress requires elaboration of the initial state and elucidation of the non-hydrodynamic degrees of freedom present early in these collisions. Here, and in the application of holographic methods to jet quenching, hybrid models incorporating holographic, perturbative, hydrodynamic or hadronic calculational methods where each is most appropriate, are needed.

- Since dissipation in the subsequent nearly perfect fluid dynamic evolution is minimal, we know that the fluid is largely transparent to fluctuations in the initial conditions. This implies that there is now a window for an experimental exploration of this system.

- Multiple hard and semi-hard interactions in hadron-hadron collisions encode information about hadron structure including quantum fluctuations. Improved experimental and theoretical knowledge of the role of multiple partonic interactions (MPI) in hadron-nucleus collisions, where they are expected to be even more important than in pp, can contribute to understanding the initial state and the earliest moments of the subsequent evolution of the produced matter.

- There has been significant progress recently in understanding how e-A scattering beyond traditional inclusive DIS measurements can place constraints on the initial nuclear wave function. Comparison of e-A to p-A can separate effects of the dynamics of partons passing through cold nuclear matter from the wave function itself. One significant challenge in this context will be to establish whether the gluons evolving from this wave function are strongly coupled, or whether they are weakly coupled but numerous as in the Color Glass Condensate framework.

## Initial conditions

Relativistic viscous hydrodynamics has emerged as a successful description for the evolution of the matter produced in relativistic heavy-ion collisions. Uncertainties in the initial conditions for hydrodynamics and their event-by-event variations are currently a major impediment to extract many QGP matter properties with a precision that matches that of the quality data measured so far. Better experimental and theoretical control over initial conditions is a necessary condition for further progress.

- By varying the collision system and energy, initial conditions can be tuned experimentally. Changing the collision system has provided important insights into the nature of the initial conditions. They help to separate collision geometry from effects due to event-by-event initial-condition fluctuations. These possibilities have so far been only partly exploited. Theoretical attempts to describe recent observations of collective dynamical behavior in p-Pb and d-Au collisions, and perhaps even in high-multiplicity pp collisions at the LHC, have highlighted the need for better understanding of event-by-event fluctuations in the initial condition, specifically for small nuclei. Collisions involving $^3$He or $^{12}$C nuclei may provide additional insights.

- There are soft physics observables, some of them requiring particle identification, that call for collecting higher event statistics over large acceptance in the future. In particular, such additional measurements will give access to finer and more differential correlation measures, with the potential to further constrain (or even over constrain) the spectrum of initial conditions. Examples include comparisons of different $p_T$-differential anisotropic flow measures for identified particles and various types of flow angle correlations that allow separating fluctuations in the magnitudes and directions of the anisotropic flows.

- Theory has made progress towards a controlled mapping of initial conditions onto hadronic and electromagnetic observables. The future emphasis of this program will be on classifying the range of conceivable fluctuations and correlations, identifying their principal components by relating them to the entire range of accessible measurements, and exploring the possibility to numerically invert the map between initial conditions and observables.

- The recent development of chiral magnetohydrodynamics (CMHD) extends the realistic dynamical modeling of heavy-ion collisions to include interesting topological effects caused by anomalous couplings between the (initially extremely strong) electromagnetic fields from the colliding nuclei, vorticity of the fluid, and non-Abelian color fields. The relative importance of the effects induced by magnetic field and vorticity is expected to change at the lower collision energies probed in the RHIC Beam Energy Scan, and at some critical energy, below which a chirally symmetric phase is no longer produced, these effects have to turn off. CMHD may thus provide an essential contribution to the quantitative interpretation of experimental observations from that campaign.

## Jets and heavy flavor

Jet quenching and heavy quark measurements probe the properties of QGP over a wide range of length scales and can thereby provide insights not accessible via other measurements. Ultimately, this includes understanding how the strongly coupled liquid at long length scales (of order $1/T$) arises, given that it is described by an asymptotically free gauge theory and so must be weakly coupled at much shorter length scales. Analogous questions are also central to other fields of physics. Initial goals of this program include the determination of average transport properties for light partons and heavy quarks moving within QGP. The achievement of both initial and longer term goals requires a combination of sensitive experimental measurements, adequate analysis observables and a self-consistent theoretical framework modeling the interaction of the probe with the medium, and the medium response to the probe, jet hadronisation, and the dynamic evolution of the medium. This framework must be capable of describing full parton showers and heavy quark propagation within strongly coupled QGP, including strongly and weakly coupled physics where each is appropriate and suitable interpolations between them as a function of momentum transfer.

- Two recent landmark developments in jet quenching are the first studies using fully reconstructed jets at the LHC, and the determination of q-hat with reduced systematic uncertainty, to date based on inclusive single hadron suppression data from RHIC and the LHC as a first step. Measurements of reconstructed jets, such as the observation of momentum imbalance by large-angle soft radiation, have provided qualitative insights, but their quantitative constraints on medium properties have yet to be fully explored.

- The lack of a closer quantitative experiment-theory connection with regards to quenching of fully reconstructed jets can be attributed, at least in part, to the complexity of the measurements and the need for an accurate match between experimental and theoretical observables. Further progress requires a contract between theorists and experimentalists that establishes observables that are both measurable and calculable with well-controlled precision.

- Promising future developments include fully reconstructed jet measurements at RHIC and multi-particle correlation measurements, together with next-generation Monte-Carlo implementations of jet transport with NLO corrections on the theory side. Future runs at the LHC and RHIC, with significantly upgraded detectors, will expand both the kinematic range and the statistical precision of jet measurements, enabling the full arsenal of techniques developed for more elementary systems to be applied to jet quenching studies.

- To match these developments, a clearer definition of goals beyond determination of an average q-hat is needed, including new observables, a program of measurements, and a self-consistent theoretical framework that permits the elucidation of both the weakly coupled short-length-scale and the strongly coupled long-length-scale physics at play in jet quenching. One set of goals includes constraining the energy, path-length and temperature dependence of the jet-medium interaction. Beyond q-hat, we should explore new observables that can be used to probe microscopic parton interactions with the medium and short-length-scale properties of the medium. We should also continue to look for jet-

- induced medium response using triggered correlations in future LHC and RHIC runs, which hold promise to measure long-length-scale properties of the medium including its viscosity and equation of state or sound velocity.

- Measurements of heavy quarks at high $p_T$ can substantially sharpen the comparison between experiment and theory, elucidating the mass-dependence and the color-charge dependence (i.e. quark vs. gluon) of energy loss and, thereby, the parton-medium interaction and transport properties. Specific considerations include the following:

    - At high $p_T$, final state effects suppress heavy quark production. The LHC data indicate a smaller suppression for B mesons than D mesons at $p_T \sim 10$ GeV/c, compatible with the mass dependence of parton energy loss expected in weakly coupled approaches. A quantitative extraction of the mass dependence requires the much better precision that will be provided by future RHIC and LHC measurements.

    - The color-charge dependence of heavy quark energy loss predicted at leading order in weakly coupled approaches is at present not revealed in experimental data on the suppression of D mesons and b-quark-initiated jets as compared to the suppression of pions and gluon-dominated jets, which in each case are compatible within current uncertainties. Models indicate, however, that these data do not at present have the precision needed to invalidate the predictions. More precise measurements of these observables, together with other measurements such as the inclusive spectra of other identified hadrons, the fragmentation function of b-quark-initiated jets at high $p_T$, and intrinsic charm with low energy fraction z in jets, promise to provide the complementary information needed to draw conclusions as to whether weakly coupled approaches suffice, or whether a strong parton-medium coupling is required to describe the data correctly.

    - The same self-consistent theoretical framework referred to above should establish what measurements of the mass dependence and the not-yet-observed color-charge dependence of energy loss imply about the quark-medium interaction and the dynamics of energy loss.

- Heavy flavor $v_2$ data at RHIC and LHC indicate that a significant fraction of the heavy quarks produced in the collisions diffuse within the strongly coupled QGP and are carried along with it as it flows. More precise measurements in upcoming RHIC and LHC runs of $v_2$ at low $p_T$ for charm quarks, and possibly also for bottom quarks, will probe the long-length-scale properties of the QGP by determining the heavy-quark diffusion constants. Advances in the first-principles computation of these quantities via lattice QCD are therefore important. These measurements, combined with measurements of heavy-flavor correlations, will also constrain the role of different energy loss mechanisms.

## Deconfinement, quarkonia and hadronisation

From AGS to LHC energies yields of hadrons with *u, d, s* valence quarks produced in collisions of heavy nuclei, yields that vary over 9 orders of magnitude, are reproduced successfully using a statistical (grand canonical) ensemble. In particular, the beam energy dependence is reproduced successfully with a smooth evolution (decrease) of the baryon chemical potential as a function of increasing collision energy and a temperature that increases from AGS to top SPS energies before, leveling off at a constant value at higher energies. This maximum temperature is empirically close to or coincident with the temperature determined by lattice QCD for the transition between quark-gluon and hadronic matter.

- At high energies (top SPS and above) this agreement is seen as a consequence of multi-particle collisions during the (cross-over) phase transition. At lower beam energies, the position of the chemical freezeout line relative to the QCD phase boundary is far from understood, implying the need for further (theoretical) work.

- At present LHC energy, there is some tension between thermal fits and data for the proton (and antiproton) yield. Results at top LHC energy as well as more precise measurements at RHIC with vertex detectors offering precise correction for a weak decay contribution, combined with dynamical modeling to quantify distortions in the chemical composition resulting from final state hadronic re-scattering that can lead to some late baryon-antibaryon annihilation, are needed to resolve the question of a possible proton anomaly.

- Explanations of the observed proton anomaly that are currently discussed include the effect of as of yet not discovered relatively low-lying baryonic states predicted by the quark model and zero temperature lattice QCD. Improving the experimental knowledge of the baryon spectrum would be very useful.

- In addition, recent analyses of recent measurements of fluctuations of conserved electric charge and baryon number at RHIC energies have shed more light on the freezeout dynamics. Extending these measurements to LHC energies and to net-strangeness fluctuations should be a priority. The analysis of such data is providing direct connection with lattice QCD calculations.

- Further input into our understanding of hadronisation and freezeout comes from measuring hadron yields and fluctuations as a function of system size. First results from p-Pb collisions have shown an intriguing trend with associated multiplicity of production yields. Such measurements can be complemented and extended by collisions of Au with d and light nuclei at RHIC and by systematically studying the system size dependence as planned in the NA61 experiment at the SPS. These experimental developments have to be accompanied by corresponding theoretical work to establish a link between hadronisation in small systems and the phase transition from quarks and gluons to hadrons for large QGP fireballs.

- Future RHIC and LHC runs will open the possibility for measuring the HQ hydrochemistry ($D_s$ and heavy-flavored baryons). This will constrain the heavy-flavor hadronization mechanisms (e.g. recombination).

- From AGS to LHC energy, the production of nuclei, anti-nuclei, hyper-nuclei follows the same statistical hadronisation picture despite the fact that their binding energies are much less than the QCD transition temperature. An interpretation might be that, after chemical freezeout, there is a purely isentropic expansion, implying that the yields of such loosely bound states are also fixed near the phase boundary. Further precision measurements would put strong constraints on the role of hadronic re-interactions after chemical freezeout and would provide the basis for a convenient and accurate computation of the yield of loosely bound states in high energy collisions. This will be relevant for hadron physics and for dark matter searches.

- The trend of the quarkonia suppression observed from SPS to LHC energies at small transverse momenta is finally putting us on the path to establish deconfinement quantitatively in the quark-gluon plasma. To establish the mechanism at work accounting for a strong charmonium suppression at SPS and RHIC and the less suppressed charmonium yields at LHC, a sequence of precise measurements is necessary.

  - Foremost is a precision measurement of the total charm production cross section by measuring the different open charm channels down to low $p_T$ to minimize extrapolation errors. The extension to full LHC energy should result in a further reduction of the measured charmonium suppression.

  - A precision measurement of the charm cross-section at RHIC energy will be necessary to demonstrate that the same mechanism is at play there. A precise measurement of the J/ψ spectrum from 0 to about 10-15 GeV/c should demonstrate the transition from statistical recombination/hadronisation of thermalized charm quarks exhibiting collective flow to the direct hard scattering component. Of course, charm quark energy loss in the plasma should play a role there, too. The measurement of J/ψ elliptic flow in the same momentum range is sensitive and should be consistent with the same physics picture. J/ψ yields may exhibit further suppression at higher momenta due to the *hot wind* scenario expected in AdS/CFT scenarios.

  - Measurement of the excited $c\bar{c}$ states ψ' and χ$_c$ should allow us to distinguish between models. In particular, such measurements should settle the question of whether or not there exist bound charmonia in the QGP.

- The prediction that the dissociation pattern of quarkonia states depends on their binding energy has been demonstrated with the measured sequential suppression of the bottomonium family. The Y(3S) state is found to be significantly more suppressed than the Y(2S) state, which is more suppressed than the Y(1S) state, reflecting the decreasing values of their binding energy. A number of measurements for bottomonia will be necessary to establish whether there also recombination of deconfined b-quarks plays a role. In particular, the question of thermalization of b-quarks needs to be addressed in the open beauty channels to be able to connect the observed suppression to Debye screening patterns in the QGP. For that precision measurements of $p_T$ spectra and the $p_T$ dependence of the nuclear modification factor in Pb-Pb and pPb collisions are needed.

## Phase diagram

A worldwide effort to survey the phase diagram of QCD is underway. Lattice QCD calculations supported by analyses of experimental data show that as QGP with low baryon chemical potential ($\mu_B$), as produced in collisions at top RHIC and LHC energies, cools and forms hadrons, this transition is a rapid but continuous crossover. Much less is known about the phase diagram of strongly interacting matter with larger $\mu_B$. Lattice calculations aiming to elucidate this regime are either indirect or very challenging or both. It is thought that the crossover may become a first order phase transition for $\mu_B$ above a critical point;

experimental discovery of such a critical point or a first order transition on the QCD phase diagram would be a landmark achievement.

- By dialing the energy of the collisions downward (to date at RHIC, by more than a factor of 25 in seven steps; via fixed-target collisions at five energies at the SPS), it is possible to study matter that freezes out at steadily increasing $\mu_B$. These experiments are scanning a region of the phase diagram with 0 < $\mu_B/T$ < 2-3. Many experimental measurements, including measurements of various flow observables and of higher moments of the event-by-event fluctuation of various hadron multiplicities, are currently being pursued with the goal of locating the critical point or the first order transition or both, if they lie within the accessible region of the phase diagram.

- Measurements made in Pb-Pb collisions at the SPS, together with those made earlier in even lower energy collisions at the AGS, show a number of rapid changes in hadronic observables such as transverse momentum spectra and particle ratios as the collision energy is varied. These measurements showed convincingly that the phase diagram can be scanned as envisioned above and suggested that collisions down to near 1/25 of top RHIC energy have enough energy to reach temperatures near or above the transition. The ongoing SPS program systematically focuses on elucidating the dependence of observables on system size. Recent discoveries in (p/d)A and pp collisions at much higher energies motivate the theoretical effort that will be needed to use SPS, RHIC and LHC data to determine below what system size, what multiplicity, and at what collision energy hydrodynamics breaks down. New experiments at the SPS in the charm and di-lepton sector are currently being considered.

- Analyses from the first phase of the RHIC beam energy scan (to date at six out of the seven energies where collisions have been done) reveal suggestive variations in fluctuation and flow observables as a function of collision energy that demand further study in a second higher statistics phase of the scan that will be made possible by the implementation of low energy electron cooling. Depending on whether the data from the recent run at √s =14.5 GeV do or do not lie on the trends defined by data at √s =7.7, 11.7 and 19.6 GeV, these data may call for measurements at further energies in this region. Regardless, present results provide strong motivation for higher statistics data at these lowest RHIC energies.

- The higher luminosity in the second phase of the RHIC beam energy scan will allow the study of matter at all but the lowest RHIC energies via di-lepton emission. At lower energies, such studies are planned at FAIR and NICA in order to bridge the energy gap between the reach of RHIC and SPS and low energy results from HADES.

- Recently, lattice QCD calculations of derivatives of the pressure at zero chemical potential with respect to $\mu_B$, $\mu_Q$ and $\mu_S$ (the latter two being the chemical potentials for electric charge and strangeness) have been used in conjunction with data on event-by-event fluctuations in heavy-ion collisions to provide a complementary way of locating where on the phase diagram the collisions freeze out. These studies should be pursued with the goal of characterizing the properties of matter in the crossover region.

- Theoretical analyses of fluctuation and flow observables have not kept pace with the quality of the new data from the RHIC beam energy scan. A concerted theoretical effort is called for.

## Collectivity and system size

The principal hallmarks of collective behavior in strongly interacting systems (radial flow, mass-dependent elliptic and triangular flow, and higher-order multi-particle correlations) have recently been observed in high-multiplicity p/dA collisions and (in part) in pp collisions. Given that these features had all been identified as unique signatures of hydrodynamic flow in heavy-ion collisions, the evident similarities between small and large systems, as well as the similarity of the results over a large span of collision energies, raises however a set of fundamental questions.

- What is the smallest (in terms of size and energy content) droplet of QGP to which a fluid dynamical description can be applied?

- When one selects high multiplicity final states in p/dA collisions or pp collisions, what features of the initial state or of the subsequent dynamics are being selected? If by selecting high multiplicity final states one is selecting collisions in which droplets of QGP are formed, how large are the protons in the initial states of the selected collisions? And, how large are the droplets of QGP that are formed?

- Is the observed collectivity in momentum space driven by the spatial structure (i.e. the pressure gradients) of the initial matter distribution created in the collision, as assumed by current hydrodynamic models?

- Is there a common way to calculate the initial conditions from first principles?

- Are there mechanisms other than hydrodynamics that can generate and quantitatively reproduce the observed collective features in these collisions?

- How does collectivity emerge as a function of system size and energy density? What are the relevant scales (time, energy, size) controlling the degree of collectivity observed in the final state?

- To which extent can a collective effect observed in a larger system be reduced to a superposition of more elementary collisions? Can this experimentally be studied by selecting events in which properties of either the smaller system or new collective effects are enhanced?

- How can we use our ability to probe different collision energies, centralities and other event characteristics and to vary the size and shape of the colliding nuclei to shed light on the mechanisms that drive the emergence of the apparent collective behavior in high-energy collisions?

- How is the onset of collective bulk dynamics in small systems correlated with hard probes of the medium, such as jet quenching and quarkonia spectroscopy?

The questions raised by the recent discovery of collectivity in what had previously been considered to be a small systems can be viewed in a larger context. Multi-hadron final states in proton-proton collisions, as well as in $e^+e^-$ annihilation to jets, have long been known to exhibit several non-trivial features that have also long been observed in heavy-ion collisions. These include the observations that the yields of produced hadrons follow thermal occupation numbers, and that HBT radii show a steady decrease with an increasing transverse mass of the particle pair. Without a crisp physical picture able to encompass both large and small systems, as well as systems that are considered to be relatively dilute like $e^+e^-$, these findings have typically been either ignored or addressed by phenomenological models. They are now more pressing.

## Bibliography

Given the generality of the arguments in this document, any referencing of specific developments would be either incomplete or of incommensurate length. Therefore, the participants have agreed to present their conclusion without referencing. Readers interested in learning more about the current state of the art in high energy nuclear collisions are referred to [1] and references therein.